\documentclass[11pt]{article}
\usepackage{graphicx}
\usepackage{epsfig}

\newcommand{\BABARPubYear}    {00}

\newcommand{\BABARProcNumber} {13}
\newcommand{\SLACPubNumber} {8696}

\def\lbabar{\mbox{{\large\sl B}\hspace{-0.4em} {\normalsize\sl A}\hspace{-0.03em}{\large\sl B}\hspace{-0.4em} {\normalsize\sl A\hspace{-0.02em}R}}}
\usepackage{relsize}
\def\babar{\mbox{\slshape B\kern-0.1em{\smaller A}\kern-0.1em
    B\kern-0.1em{\smaller A\kern-0.2em R}}}

\def\piz   {\ensuremath{\pi^0}}

\def\pip   {\ensuremath{\pi^+}}
\def\pim   {\ensuremath{\pi^-}}

\def\Kbar  {\kern 0.2em\overline{\kern -0.2em K}{}}

\def\Kp    {\ensuremath{K^+}}

\def\KS    {\ensuremath{K^0_{\scriptscriptstyle S}}} 
 
\def\Kstarz  {\ensuremath{K^{*0}}}

\def\Kzb   {\ensuremath{\Kbar^0}}
\def\KzKzb {\ensuremath{K^0 \kern -0.16em \Kzb}}

\def\Dbar  {\kern 0.2em\overline{\kern -0.2em D}{}}

\def\Dzb   {\ensuremath{\Dbar^0}}
\def\DzDzb {\ensuremath{D^0 {\kern -0.16em \Dzb}}}

\def\Bz    {\ensuremath{B^0}}
\def\B     {\ensuremath{B}}
\def\Bbar  {\kern 0.18em\overline{\kern -0.18em B}{}}

\def\Bzb   {\ensuremath{\Bbar^0}}
\def\Bu    {\ensuremath{B^+}}

\def\BB    {\ensuremath{B\Bbar}} 
\def\BzBzb {\ensuremath{B^0 {\kern -0.16em \Bzb}}}

\mathchardef\Upsilon="7107
\def\Y#1S{\ensuremath{\Upsilon{(#1S)}}}

\def\FourS {\Y4S}

\mathchardef\Deltares="7101
\mathchardef\Xi="7104
\mathchardef\Lambda="7103
\mathchardef\Sigma="7106
\mathchardef\Omega="710A
\def\Deltabar   {\kern 0.25em\overline{\kern -0.25em \Deltares}{}}
\def\Lbar {\kern 0.2em\overline{\kern -0.2em\Lambda\kern 0.05em}\kern-0.05em{}}
\def\Sigbar{\kern 0.2em\overline{\kern -0.2em \Sigma}{}}
\def\Xibar{\kern 0.2em\overline{\kern -0.2em \Xi}{}}
\def\Obar{\kern 0.2em\overline{\kern -0.2em \Omega}{}}
\def\Nbar{\kern 0.2em\overline{\kern -0.2em N}{}}
\def\Xbar{\kern 0.2em\overline{\kern -0.2em X}{}}

\def\Bztokpi    {\ensuremath{B^0 \to K^{\pm}\pi^{\mp}}}

\def\mes        {\mbox{$m_{\rm ES}$}}

\def\ev   {\ensuremath{\rm \,e\kern -0.08em V}}
\def\kev  {\ensuremath{\rm \,ke\kern -0.08em V}} 
\def\mev  {\ensuremath{\rm \,Me\kern -0.08em V}} 
\def\gev  {\ensuremath{\rm \,Ge\kern -0.08em V}} 
\def\gevc {\ensuremath{{\rm \,Ge\kern -0.08em V\!/}c}} 
\def\tev  {\ensuremath{\rm \,Te\kern -0.08em V}}
\def\mevc {\ensuremath{{\rm \,Me\kern -0.08em V\!/}c}} 
\def\gevcc{\ensuremath{{\rm \,Ge\kern -0.08em V\!/}c^2}} 
\def\mevcc{\ensuremath{{\rm \,Me\kern -0.08em V\!/}c^2}}

\def\invfb   {\ensuremath{\mbox{\,fb}^{-1}}}
\def\mus  {\ensuremath{\rm \,\mus}}

\def\mus        {\ensuremath{\,\mu{\rm s}}}    
\def\gsim{{~\raise.15em\hbox{$>$}\kern-.85em
          \lower.35em\hbox{$\sim$}~}}
\def\lsim{{~\raise.15em\hbox{$<$}\kern-.85em
          \lower.35em\hbox{$\sim$}~}}

\def\CP                 {\ensuremath{C\!P}}

\def\to                 {\ensuremath{\rightarrow}}

\def\pep2{PEP-II}

\newcommand{\etapr}{\ensuremath{\eta^{\prime}}}

\newcommand{\dedx}{\ensuremath{\mathrm{d}\hspace{-0.1em}E/\mathrm{d}x}}

\newcommand{\eqref}[1]{Eq.~(\ref{eq:#1})}

\newcommand{\plb}       [1]  {{Phys.\ Lett.\ B~{\bf #1}}}   

\def\jetset74   {\mbox{\tt Jetset \hspace{-0.5em}7.\hspace{-0.2em}4}}

\newcommand{\dzpi}{$\Bu \to \Dzb \pip, \Dzb \to \Kp \pim$}

\newcommand{\kstpi}{$\Bu \to \Kstarz \pip$}
\newcommand{\rhok}{$\Bu \to \rho^0 \Kp$}
\newcommand{\rhozpipm}{\mbox{$\Bu \to \rho^0 \pip$}}
\newcommand{\ppp}{$\Bu \to \pip \pim \pip$}
\newcommand{\kpp}{$\Bu \to \Kp \pim \pip$}

\newcommand{\rhoppim}{\mbox{$\Bz \to \rho^{\mp} \pi^{\pm}$}}

\newcommand{\etapKp}{\mbox{$\Bu \to \etapr \Kp$}}

\newcommand{\etapKs}{\mbox{$\Bz \to \etapr \KS$}}

\newcommand{\omegaKs}{\mbox{$\Bz \to \omega \KS$}}

\newcommand{\de}{\ensuremath{\Delta E}}

\long\def\inst#1{\par\nobreak\kern 4pt\nobreak
    {\it #1}\par\vskip 10pt plus 3pt minus 3pt}

\begin{document}
{\pagestyle{empty}

\begin{flushright}
SLAC-PUB-\SLACPubNumber \\
\babar-PROC-\BABARPubYear/\BABARProcNumber \\
August, 2000 \\
\end{flushright}

\par\vskip 4cm

\begin{center}
\Large \bf 
 Studies of Charmless Two-Body, Quasi-Two-Body and Three-Body \B\ Decays
\end{center}
\bigskip

\begin{center}
\large 
Theresa J. Champion\\
University of Birmingham, Edgbaston,\\ Birmingham, England.\\
E-mail: tjc@SLAC.Stanford.edu \\
(for the \lbabar\ Collaboration)
\end{center}
\bigskip \bigskip

\begin{center}
\large \bf Abstract
\end{center}
Preliminary results are presented on a search for several exclusive
charmless hadronic $B$ decays, from data 
collected by the \babar\ detector near the \FourS\ resonance. 
These include two-body decay modes $h^{\pm}h^{\mp}$, three-body decay 
modes with final states 
$h^{\pm}h^{\mp}h^{\pm}$ and $h^{\pm}h^{\mp}\pi^0$, and quasi-two-body 
decay modes with final states $X^0 h$ and $X^0 \KS$, where $h = \pi$ or 
$K$ and $X^0 = \etapr$ or $\omega$.
The measurement of branching fractions for four decay modes, and upper 
limits for nine modes are presented.

\vfill
\begin{center}
Contributed to the Proceedings of the 30$^{th}$ International 
Conference on High Energy Physics, \\
7/27/2000---8/2/2000, Osaka, Japan
\end{center}

\vspace{1.0cm}
\begin{center}
{\em Stanford Linear Accelerator Center, Stanford University, 
Stanford, CA 94309} \\ \vspace{0.1cm}\hrule\vspace{0.1cm}
Work supported in part by Department of Energy contract DE-AC03-76SF00515.
\end{center}

\setlength\columnsep{0.20truein}
\twocolumn
\def\sloppy{\tolerance=100000\hfuzz=\maxdimen\vfuzz=\maxdimen}
\sloppy
\vbadness=12000
\hbadness=12000
\flushbottom
\def\figurebox#1#2#3{%
  	\def\arg{#3}%
  	\ifx\arg\empty
  	{\hfill\vbox{\hsize#2\hrule\hbox to #2{\vrule\hfill\vbox to #1{\hsize#2\vfill}\vrule}\hrule}\hfill}%
  	\else
   	{\hfill\epsfbox{#3}\hfill}%
  	\fi}

\section{Introduction}

Charmless hadronic decays of $B$ mesons provide rich opportunities
for exploring a number of \CP\ violation phenomena.
In particular, several of these decay modes offer the future possibility 
of measuring directly the CKM angle $\alpha$ of the Standard Model
\cite{ref:physbook}.

Preliminary searches have been carried out for a number of charmless 
hadronic $B$ decays using the initial \babar\ data sample. The data 
consist of 7.7\invfb\ taken at the \FourS\ resonance, and 1.2\invfb\ 
taken below the \BB\ threshold. The number of \BB\ events has been determined 
from hadronic event selection\cite{ref:conf15} to be $8.46 \pm 0.14 \times 10^6$.


For all decay modes a simple cut-based analysis has been performed, and additionally 
for the two-body modes a global likelihood fit has been carried out.
A ``blind'' analysis methodology has been adopted throughout, so that the signal 
region for each mode remained hidden until all decisions concerning event 
selection had been taken. The signal region has been represented using the two 
variables $\mes = \sqrt{(\sqrt{s}/2)^2 - P_B^{*2}}$ and $E_B^* - \sqrt{s}/2$, where 
$E_B^*$ and $P_B^{*}$ are the energy and 3-momentum of the $B$ in the cms. 
Fig.\ref{fig:q2body} shows the distribution of events in the 
\mes\ -- \de\ plane for the mode \etapKp.

In order to facilitate understanding of the data, use was made of 
suitable calibration modes \dzpi\ and $\Dzb \to \Kp \pim \piz$. These 
have similar final state kinematics to the modes of interest, but with 
$\sim \times 10$ higher branching ratios.
\vspace{-1mm}
\section{Event Selection}
The most significant issues for event selection, common to all the rare 
charmless decay modes, are effective kaon identification and the suppression 
of continuum background. Additional features of individual modes, such as 
the reconstructed mass and helicity angle distributions of intermediate 
resonances, have been used for selection where appropriate (for details see 
[2],[3]).

\begin{figure}
\epsfxsize180pt
\figurebox{120pt}{160pt}{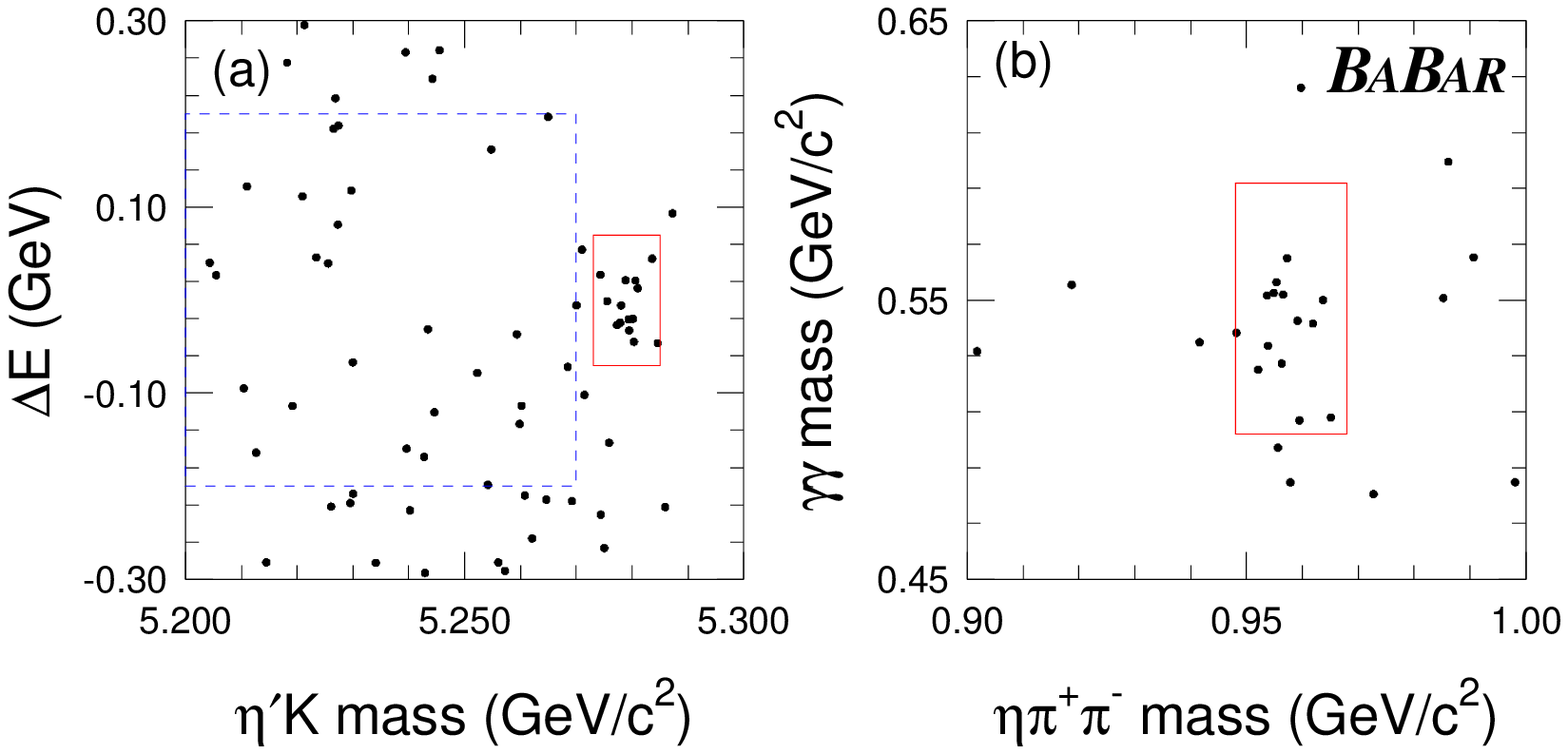}
\caption{Kinematics of \etapKp.}
\label{fig:q2body}
\end{figure}

\vspace{-1mm}
\subsection{Kaon Identification}
Excellent kaon identification is essential for all the decay modes 
of interest. The primary system in \babar\ for $K/\pi$ discrimination is 
the DIRC (Detector of Internally Reflected Cherenkov light), which achieves 
separation at $>2\sigma$ for momenta up to 4\gev \cite{ref:conf17}.
Further information for particle identification is provided by \dedx\ 
from the Drift Chamber and the Silicon Vertex Tracker.

\begin{figure}
\epsfxsize180pt
\figurebox{120pt}{160pt}{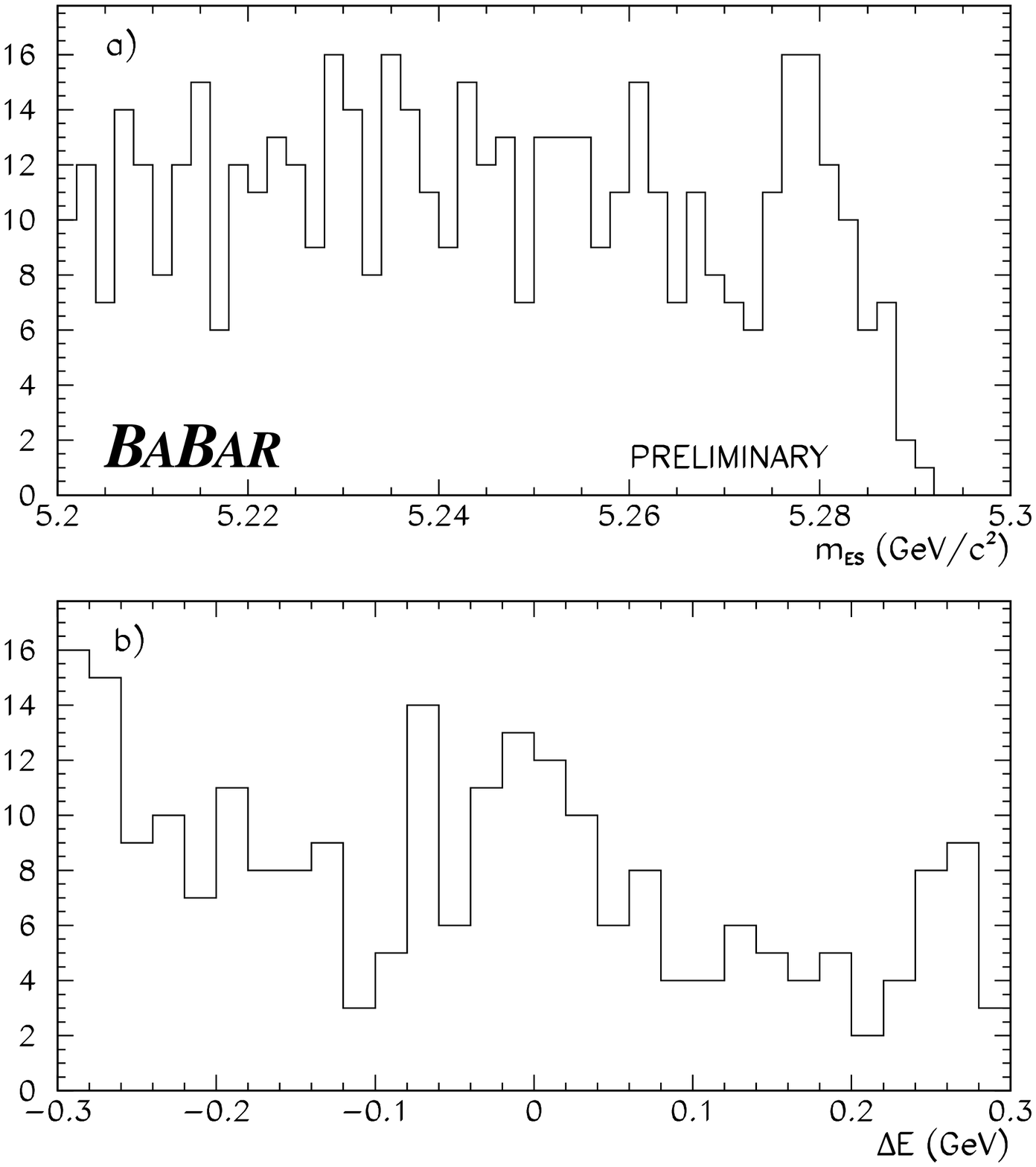}
\caption{(a) \mes\ and (b) \de\ distributions for \rhoppim.}
\label{fig:3body}
\end{figure}
\vspace{-1mm}
\subsection{Background Characterisation and Suppression}
All rare charmless $B$ decays suffer from high levels of background 
from continuum events. This background can be substantially reduced 
by exploiting the differences in topology between \BB\ events and 
continuum events. In the cms, a \BB\ event is approximately 
isotropic, and there is no correlation between the decay topologies of the 
two $B$s. A continuum event, however, exhibits a two-jet structure, 
so that the directions of the decay products are highly correlated.
A number of event shape variables have been used in the analyses to 
discriminate signal events from continuum background. 
Following background discrimination cuts, a significant amount 
of background remained. This has been estimated using 
both on- and off-resonance data. In the on-resonance case, the background has been 
characterised by fitting the \mes\ distribution to an ARGUS function\cite{ref:Argus} 
using the sidebands, and extrapolating to the signal region. In the off-resonance case, 
the number of events in the sidebands and signal region were counted directly.

\vspace{-1mm}
\section{Analysis}
\begin{figure}
\epsfxsize160pt
\figurebox{120pt}{160pt}{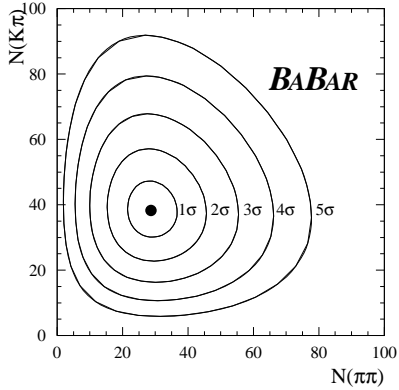}
\caption{Likelihood contours for $\pi ^\pm \pi ^\mp$ and $K^\pm \pi >^\mp$.}
\label{fig:2body}
\end{figure}

For the simple counting analyses, cut optimisation was performed with respect 
to the predicted sensitivity of a measurement, using the expected signal yield 
and the estimated efficiency. Fig.\ref{fig:3body} shows the distributions of 
\mes\ and \de\ for the mode \rhoppim.

For the two-body modes, a global likelihood fit was carried out. A likelihood 
function was constructed to determine the signal and background yields, using
 the parameters \mes, \de, the Fisher Discriminant formed from a set of event 
shape variables, and the Cherenkov angles of the two decay products. The 
probability density functions used in the fit were obtained from studies of 
the data where possible. The likelihood contours for $\pi ^\pm \pi ^\mp$ and 
$K^\pm \pi >^\mp$ are shown in Fig.\ref{fig:2body}.

The results of searches for 13 decay modes are presented in Table \ref{tab:results}.

\begin{table*}[t]
\caption{Summary of Results.\label{tab:results}}
\begin{tabular}{|p{162pt}|p{47pt}|l|l|}  
\hline
\raisebox{0pt}[16pt][6pt]
{Decay mode} 
&yield
&\babar\ BR$/10^{-6}$
&CLEO\cite{ref:PDG2k} BR$/10^{-6}$
\\[6pt]
\hline \hline
\raisebox{0pt}[16pt][6pt]
{$B^0 \to \pi^{\pm}\pi^{\mp}$ }
&$25 \pm 8$
&$9.3^{+2.6+1.2}_{-2.3-1.4}$
&$4.3^{+1.6}_{-1.4}\pm 0.5$
\\[6pt] 
\hline
\raisebox{0pt}[16pt][6pt]
{\Bztokpi }
&$26 \pm 8$
&$12.5^{+3.0+1.3}_{-2.6-1.7}$
&$17.2^{+2.5}_{-2.4}\pm 1.2$
\\[6pt] 
\hline
\raisebox{0pt}[16pt][6pt]
{$B^0 \to K^{\pm}K^{\mp}$ }
&$1 \pm 4$
&$< 6.6$
&$< 1.9$
\\[6pt] 
\hline\hline
\raisebox{0pt}[16pt][6pt]
{$\B^+ \to \omega h^+$ }
&$6 \pm 4$
&$< 24$
&$< 14.3 \pm 3.6$
\\[6pt] 
\hline
\raisebox{0pt}[16pt][6pt]
{\omegaKs }
& 0
&$< 14$
&$< 21$
\\[6pt] 
\hline
\raisebox{0pt}[16pt][6pt]
{\etapKp }
&$12 \pm 4$
&$62 \pm 18 \pm 8$
&$80 \pm 10$
\\[6pt] 
\hline
\raisebox{0pt}[16pt][6pt]
{\etapKs }
&$1 \pm 1$
&$< 112$
&$< 89 \pm 18$
\\[6pt] 
\hline
\hline
\raisebox{0pt}[16pt][6pt]
{\kstpi }
&$10 \pm 5$
&$< 28$
&$< 16$
\\[6pt] 
\hline
\raisebox{0pt}[16pt][6pt]
{\rhok }
&$11 \pm 5$
&$< 29$
&$< 17$
\\[6pt] 
\hline
\raisebox{0pt}[16pt][6pt]
{\kpp }
&$19 \pm 6$
&$< 66$
&$< 28$
\\[6pt] 
\hline
\raisebox{0pt}[16pt][6pt]
{\rhozpipm }
&$25 \pm 8$
&$< 39$
&$< 10.4 \pm 3.4$
\\[6pt] 
\hline
\raisebox{0pt}[16pt][6pt]
{\ppp }
&$5 \pm 6$
&$< 22$
&$< 41$
\\[6pt] 
\hline
\raisebox{0pt}[16pt][6pt]
{\rhoppim }
&$36 \pm 10$
&$< 48.5 \pm 13.4 ^{+5.8}_{-5.2}$
&$< 27.6 \pm 8.4$
\\[6pt] 

\hline
\end{tabular}
\end{table*}

\end{document}